\documentstyle[11pt,aasms4]{article}

\def\alwaysmath#1{\ifmmode{#1}\else{$#1$}\fi}

\righthead{Paltrinieri, et al.}
\lefthead{VLT OBSERVATIONS OF GLOBULAR CLUSTER NGC\,6712}

\slugcomment{Astronomical Journal in Press}
\begin{document}

\title
{VLT Observations of the Peculiar Globular Cluster NGC\,6712, III: 
 The Evolved Stellar Population
\altaffilmark{1} 
}

\author{
Barbara~Paltrinieri\altaffilmark{2}, 
Francesco~R.~Ferraro\altaffilmark{3},   
Francesco Paresce\altaffilmark{4},  
Guido De Marchi\altaffilmark{4,5,6}} 
 
\altaffiltext{1}{Based on data taken at the VLT, European Southern
Observatory, Chile, as part of the ESO observing programme
and 63.L-0423(A).}
\altaffiltext{2}{Istituto di Astronomia --- Universit\'a La Sapienza,
Via G.M. Lancisi 29, I--00161 Roma, Italy;
barbara@coma.mporzio.astro.it}
\altaffiltext{3}{Osservatorio Astronomico di Bologna, via Ranzani 1,
I--40126 Bologna, Italy; ferraro@bo.astro.it}
\altaffiltext{4}{European Southern Observatory, Karl Schwarzschild
Strasse 2, D--85748 Garching bei M\"unchen, Germany;
fparesce@eso.org}
\altaffiltext{5}{Space Telescope Science Institute, 3700 San Martin
Drive, Baltimore, MD 21218, USA; demarchi@stsci.edu}
\altaffiltext{6}{Affiliated to the Astrophysics Division, Space
Science Department of ESA}

\begin{abstract}

We present extensive UBVR photometry of the Galactic globular cluster
(GGC) NGC\,6712 obtained with the ESO Very Large Telescope (VLT) which
reach down to two magnitudes below the main sequence turn-off and
allows us for the first time to determine the age of this cluster. By
using the apparent luminosity of the zero age horizontal branch (ZAHB),
$V_{ZAHB}=16.32\pm0.05$ and the stellar main sequence (MS) turn--off
(TO) magnitude $V_{TO}=19.82\pm0.10$, we obtain $\Delta
V_{TO}^{HB}=3.5\pm0.1$ (a value fully compatible with that derived for
other clusters) which suggests that, at an age of $\sim 12$ Gyr,
NGC\,6712 is coeval with other GGC of similar metallicity.

We derive interstellar reddening by comparing the position and
morphology of the red giant branch (RGB) with a wide variety of
reference clusters and find $E(B-V)=0.33\pm0.05$, a value significantly
lower than had been determined previously. Assuming this value for  the
reddening, we  determine a true distance modulus of $(m-M)_0 = 14.55$,
corresponding to a distance of $\sim 8$ kpc.

We find a population of 108 candidate blue straggler stars (BSS),
surprisingly large when compared with the typical BSS content of other
low concentration clusters. Moreover, we detect a very bright blue star
in the core of NGC\,6712 that might be a post-AGB star. These results,
combined with those already presented in two companion papers, strongly
support the hypothesis that NGC\,6712 was, at some early epoch of its
history, much more massive and concentrated. The continued interaction
with the bulge and the disk of the Galaxy has driven it toward
dissolution, and what we now observe is nothing but the {\it remnant}
core of a cluster that once was probably one of the most massive  in
the Galaxy.

\end{abstract}


\section{Introduction}

NGC\,6712 is a low-concentration ($c=0.9$, $Log \rho_0 \sim 3$;,
Djorgovski \& Meylan 1993), intermediate-high  metallicity ($[Fe/H]=-1.01$
Zinn 1985) globular cluster, which has, without doubt, some peculiar
characteristics.  First, it is the lowest density GGC containing a low
mass X-ray binary (LMXB - namely $1RXS J185304.8-084217$; Voges et al.
1999).  LMXB are thought to form via tidal capture in high density
clusters, thus the presence of such an object in the core of a loose
cluster is somewhat surprising.  Second, its orbit, as computed by
Dauphole et al. (1996), suggests that it is experiencing a severe
interaction with the Galaxy due its numerous passages through the disk
and bulge.

Within this intriguing scenario, we have undertaken a detailed study of
the global stellar population in NGC\,6712, using the ESO {\it Very
Large Telescope}. This is the third paper in a series dedicated to this
cluster.

Ferraro et al.2000a (hereafter {\it Paper I}), besides confirming the
optical counterpart to the LMXB, discovered the presence of an
additional interacting binary candidate (showing strong UV and
$H\alpha$ excess) in the core of NGC\,6712, located a few arcsec from
the LMXB.

Deep $V$- and $R$-band observations of the lower MS, presented by
Andreuzzi et al. 2001 (hereafter {\it Paper II}) confirmed the previous
findings by De Marchi et al. (1999) that the mass function of this
cluster has been severely depleted of lower mass stars, probably
stripped away by the tidal force of the Galaxy.

These results add further support to the scenario that NGC\,6712 is
only a pale remnant of a once much more massive cluster.  Takahashi \&
Portegies Zwart (2000), using extensive N-body simulations, have
recently shown that the inverted mass function observed by De Marchi et
al. (1999) is a clear signature of significant mass evaporation since
it only appears in clusters which have lost most (up to $\sim 99\%$) of
their initial mass.  If this scenario is confirmed, with a current
mass of $\sim 10^5  M_{\odot}$ (Pryor \& Meylan 1993) NGC\,6712 might
have been one of the most massive clusters ever formed in the Galaxy
with $M_{inital} \sim 10^7  M_{\odot}$.

In this paper, we make use of a large data-set obtained with the VLT in
order to present and discuss the overall morphology and the main
characteristics of the CMD for stars brighter than the MS TO, looking
for any additional signature of the stormy past of this cluster.

The plan of the paper is as follows. We present the observations and
data analysis Section 2 while the overall properties of the resulting
colour--magnitude diagramme (CMD) are described in Section 3.
Sections 4, 5, 6, 7 are devoted each to the description of one of the
evolutionary sequences (horizontal branch - HB, RGB, blue straggler
stars - BSS, and blue objects, respectively).  A discussion of the
cluster age and distance is given in Section 8.  Section 9 summarises
the main results of this work.

\section{Observations and data reductions}

\subsection{Observations}

A large set of frames covering the central region of NGC\,6712 was
obtained in service mode during June 1999 at the {\it ANTU} Very Large
Telescope (VLT-UT1) at ESO on Cerro Paranal (Chile) using the FORS1
camera equipped with a $2048\times2048$ pixel$^2$ CCD array).  The data
were obtained using two different levels of resolution:  {\it (i)} high
resolution (HR), with a plate-scale of $0\farcs1/pixel$ and a field of
view of $3\farcm4 \times 3\farcm4$; and {\it (ii)} standard  resolution
(SR) with a   plate-scale of $0\farcs2/pixel$ and a field of view of
$6\farcm8 \times 6\farcm8$.  HR frames were roughly centered on the
cluster centre, while the four partially overlapping SR fields (one for
each quadrant, Q1, Q2, Q3, Q4, respectively) were roughly centered at
$3\farcm5 $ for the cluster centre(see Figure 1).  This configuration
allowed us to observe the innermost regions of the cluster with the
most appropriate resolution and to cover a large area ($\sim 13\farcm
\times 13\farcm$) around the cluster.

The data consist of five 10\,s $B$-, $V$-, $R$-band exposures, five
120\,s $U$-band exposures for each HR and SR field.  An additional
700\,s $H\alpha$ exposure was obtained only in  the HR field.

\subsection{Magnitudes, colour and positions}

All the reductions were carried out using ROMAFOT (Buonanno et al.
1983, Buonanno \& Iannicola 1986), a package specifically developed to
perform accurate photometry in crowded fields. Specific subroutines of
this package have been used to detect stars in each frame, in
particular independent searches have
 been performed in the best blue ($U,B$) and in the best red ($V,R$)
images, in order to properly optimize the search for blue and red
objects.  The {\it blue} and the {\it red} masks built in this way and
containing the position of  the stars have been then adapted to each
{\it colour-group} image of the same field (the red mask to all the
$V$- and $R$-band images and the blue mask for the $U$ and $B$ frames)
and the standard PSF-fitting procedure performed.  The output
instrumental magnitudes in the same band were then averaged by properly
weighing  the
photometric quality of individual frames to yield the adopted {\it
u,b,v,r} mean instrumental magnitudes.  The  data-sets were then
matched  together and a final catalogue with stellar coordinates and
the average instrumental magnitude in each filter has been compiled for
all the objects identified in each of the FORS1 fields.

Stars in overlapping areas (see Figure 1) were used to transform the X
and Y coordinates of each field to a common coordinate system.
Multiple measurements of the objects common to adjacent SR fields were
averaged, while for objects common to HR and a SR field, the HR
magnitudes were taken owing to the higher resolution and quality of
these images.  The result of this procedure is a homogeneous set of
instrumental magnitudes, colours and positions for all the stars in our
frames.

\subsection{Calibration}

The final instrumental magnitudes were then transformed into the
standard Johnson photometric system by using ten photometric standard
stars in selected areas PG1528, PG2213, PG2331 (Landolt 1992).
Standard aperture photometry was performed on these stars.  The
calibration curves linking the instrumental aperture photometry to the
Johnson magnitudes are plotted in Figure 2.  A subset of the brightest,
unsaturated and most isolated stars was used to link the {\it
instrumental}  magnitudes obtained with PSF fitting to those derived
with aperture photometry.

$H\alpha$ magnitudes were treated and calibrated as $R$-band images,
except for an offset of $\sim 4$ mag that was applied so as to account
for the lower filter efficiency.

The calibrated data can be compared   with the photographic $V$- and
$B$-band data of Cudworth (1988, hereafter C88):  the residuals in
magnitude computed using a sample of $\sim 100$ stars in common with
C88 turn out to be very small in both the B and V filters
($V-V_{C88}=0.00$ and $B-B_{C88}=-0.03$) showing an excellent agreement
between the photometric zero-points used here and those of C88.

\subsection{Photometric errors}

Since we have 5 repeated exposures through each filter, the internal
accuracy of our photometry has been estimated from the rms
frame-to-frame scatter of the instrumental magnitudes.  We computed this
quantity for each individual star according to the formula given in
Ferraro et al. (1991).  The rms values obtained for the $U$, $B$, $V$
and $R$ frames are plotted in Figure 3 as a function of the mean final
calibrated magnitude adopted for each star measured in the HR field.
The mean photometric errors are very small ($\sigma <0.05$) over the
whole magnitude range covered by our observations in all  filters.

\section{Colour-Magnitude Diagrammes: overall structure}

As result of the procedure described in the previous section,  a
catalogue with magnitudes, colour and positions for a sample of 30,000
stars has been obtained in the  region covered by our 
observations\footnote{The electronic version of the table 
listing magnitudes and positions for the entire sample
is available upon request from the second author (e-mail: ferraro@bo.astro.it)}
(see Figure 1).  Figure 4 shows the CMD corresponding to stars in  regions
located at different distances from the cluster centre.  In particular,
in {\it  panel (a)}  all the stars lying in the HR field are plotted
while the other three panels ({\it (b),(c)} and {\it (d)},
respectively) show the stars located in annuli at increasing distances
from the cluster centre.  Here the centre of the cluster has been
assumed to be the one defined in {\it Paper I}.

The inspection of the CMD presented in Figure 4 shows that field
contamination is particularly severe around NGC\,6712. This  was to be
expected, since the cluster is located behind the Scutum stellar cloud
which is one of the highest surface brightness regions in the Milky
Way.

The overall characteristics of the CMD in Figure 4 can be summarised as
follows:

\begin{itemize}

\item{(1)} The population sampled by the HR field is  dominated by
cluster stars:  all sequences in the CMD, namely the RGB, HB, sub-giant
branch (SGB), TO region, BSS sequence, are well defined and populated.
The field component, however, is also  visible (note the field MS
crossing the diagram from $B-V\sim 0.4, V\sim14.$ down to $B-V\sim
1.2, V\sim 20.5$);

\item{(2)} as expected, at increasing distance from the cluster centre,
the field contribution rapidly increases with respect to the cluster
population;

\item{(3)} although in the region $150''<r<360''$ the field population
dominates the overall morphology of the CMD, some features of the
cluster population are still visible, such as a few blue HB stars at
$18<V<17$ and $(B-V)<0.5$ (see Figure 4{\it b} and 4{\it c}).

\item{(4)} blue HB stars are not visible in the CMD at distances
$r>360''$ (see Figure 4 {\it d}).

\end{itemize}

\subsection{ Field decontamination}
  
Some features in Figure 4 show that the cluster population is still
present in the region $150''<r<360''$, even if  evolutionary sequences
are not clearly defined in the CMD.  On the other hand the absence of
blue HB stars for $r>360''$ suggests that the cluster population
becomes negligible beyond this distance.

This indication is fully confirmed by the investigation of MS stars
presented in {\it Paper II}, where we compared the MS cluster
population at different distances from the cluster centre with that
obtained in a reference field taken at much larger distance ($\sim
42^{\prime}$).  From this comparison   we concluded that the  stellar
population at $r>5'$ from the cluster centre is fully consistent with
that observed in the reference field, i.e. it is made almost
exclusively of field stars.  In Figure 5 we present the projected
density distribution ({\it filled circles}) for the evolved population
of the cluster. All stars  with $V<20$ plotted in Figure 4 have been
counted. The stellar density has been scaled to match the  density
profile of MS stars  with $19.5<R<20.5$ (corresponding to $\sim 0.75
M_{\odot}$) and $r>2'$ (plotted as {\it large open squares} in Figure 5
-- see also Figure 8 in {\it Paper II}).  The figure clearly shows a
plateau in the stellar density for $r>4.2'$, which is fully consistent
with the density of $ 0.75 M_{\odot}$-mass star measured in the control
field at $42'$ from the cluster centre (indicated as the {\it dotted
horizontal line} in Figure 5). This confirms that the cluster
population becomes indeed negligible for $r>5'$.
   
On the basis of these considerations, we conservatively assumed that
the sample at  $r>360''$ can be regarded as field population (see Figure
4(d)) and used it in order to decontaminate the CMD.To be sure, exact
discrimination between cluster and field members is not possible on the
basis of the CMD comparison alone. However, in order to study the
global characteristics of the stellar population of NGC\,6712, we
adopted a statistical approach, following the method described by
Mighell et al. (1996) and successfully used by others as well (see for
example Bellazzini, Ferraro \& Buonanno, 1999, Testa et al. 1999).  In
short, for each star in the {\it contaminated} CMD, the probability of
membership is computed by comparing the number of stars lying within a
cell (of assigned size) in the {\it contaminated} CMD and in the {\it
field} CMD.

The decontamination procedure has been applied to each of the CMD
plotted in Figure 4 {\it a, b, c}. Only in the case of the HR field,
however, does the  procedure yield acceptable results.  Figure 6 shows
the statistically decontaminated CMD of the HR field.  Here the result
is quite good as the statistical decontamination successfully removes
most of the field stars from the HR CMD and the cluster sequences
appear well defined. For this reason, we conservatively decided to
limit the analysis of the stellar population of NGC\,6712 to this
region.

\subsection{Cluster population: ridge lines}

Figure 5 shows that all the sequences in the CMD are  well defined and
populated. Particularly noteworthy is the large population of BSS (see
Section 6) which has not been affected by the field decontamination
procedure as the field MS appears to be significantly redder than the
BSS sequence (see Figure 4{\it a} and 4{\it d}), mostly because of the
higher metallicity of the field population.

The mean ridge lines of the main branches (RGB, SGB) in the CMD plotted
in Figure 6 were determined following an iterative process. A first
rough selection of the candidate RGB stars was performed by eye,
removing the HB and part of the AGB stars. Then, the sample was divided
into $0.4$ V-mag wide bins and the median colour and magnitude computed
in each bin. Stars more distant than $2\sigma$ from the median point
were rejected and the median and $\sigma$ recomputed until no more
outliers were found, The median points were then interpolated through a
spline over an interval of typically 3 magnitudes. The resulting ridge
line for the RGB-SGB and TO-MS region is listed in Table 1.

\section{The HB morphology and the ZAHB level}

Figure 7 plots HB stars from the decontaminated sample of Figure 6.
The bulk of the HB stars populates the red side of the instability
strip, as expected since NGC\,6712 is a cluster of intermediate-high
metallicity ($[Fe/H]_{Z85}=-1.01$).
 
Our CCD photometry does not have the proper time-coverage to search for
variability, for this reason we identified the RR-Lyrae variables 
lying in the FORS1-HR field of view  
from previous studies.
The 7 known RR Lyrae variables (namely V1, V5, V6, V12, V18, V19, V20)
identified in  our field are plotted as large filled triangles
in Figure 7. Their position in the CMD clearly shows that we observed
these variables at random phases of their pulsation cycle. Since our
observations were not homogeneously spread over the pulsation period,
we are unable to derive meaningful average magnitudes and colours for
these objects.

In order to more quantitatively describe the morphology of the HB, we
computed the parameter $HBM=(B-R)/(B+V+R)$ (Lee et al. 1990, but see
Fusi Pecci et al. 1992, Buonanno et al. 1997), where $B$ and $R$ are
the number of stars bluer and redder than the RR Lyrae gap,
respectively, and $V$ is the number of RR Lyrae variables.  Using our
sample, this parameter takes on the value
$HBM=(23-127)/(23+7+127)=-0.66\pm0.12$, a figure typical of clusters of
intermediate-high metallicity which confirms that  the HB morphology of
NGC\,6712 closely resembles that of NGC\,6171 (Ferraro et al. 1991), a
cluster with similar metallicity.

To provide a characterisation of the extreme blue HB population with
respect to the red one, we also computed  Buonanno's (1993) index
$P_{HB}=(B2-R)/(B+V+R)$, where $B2$ is the number of the blue HB stars
bluer than $(B-V)_0=-0.02$. This parameter is, of course,
reddening-dependent: assuming $E(B-V)=0.33$, as explained in Section
5.2,we obtain $P_{HB}=-0.78\pm0.14$.

Ferraro et al. (1999a, hereafter F99) presented a new procedure for the
determination of the level of the ZAHB in the observed (V,B-V)-CMD.
This procedure uses synthetic HB models   to reproduce the observed HB
morphology. Once a proper agreement between the observed and the
synthetic HB has been obtained, the value of $V_{ZAHB}$ can be derived
from  the  model which has been used to construct the synthetic HB.

F99 derived $V_{ZAHB}=16.32\pm0.05$ for NGC\,6712 adopting the C88
photometry. As shown in Section 2.3,  the C88 photometry nicely agrees
with that presented in this paper. Thus, in the following, we adopt
this  value  for the ZAHB level. The ZAHB is shown as a solid line in
Figure 6; as expected, it is consistent with  the lower boundary of the
magnitude distribution of the red HB clump.

\section{The RGB: metallicity and reddening}

Once properly calibrated, the shape and position of the RGB 
in the CMD can be used to derive photometric estimates 
of the cluster metallicity.

\subsection{Metallicity scales}

The most widely used metallicity scale for GGC is that proposed by Zinn
and collaborators in the 1980s (Zinn \& West 1984, Zinn 1985) and
essentially based on integrated light indexes and low resolution
spectra. There is now a growing body of reasons (see Carretta \&
Gratton 1997 - herafter CG97 and Rutledge et al. 1997) suggesting a
revision of the {\it classical} Zinn scale, according to the results of
the latest high-resolution spectral data which have the advantage of
measuring directly Iron absorption lines.  In the following, we adopt
the CG97 metallicity scale ($[Fe/H]_{CG97}$).

Moreover, there is much observational evidence that {$\alpha$}-elements
(such as Si and Ca) are often enhanced with respect to Iron for
Population II stars and thus the knowledge of the Iron abundance alone
in not sufficient to correctly evaluate the {\it global} metal content
in each cluster.  For this reason, the {\it global} metal index
($[M/H]$) (including also the   {$\alpha$}-element enhancement) has
been defined (see Salaris, Chieffi \& Straniero, 1993). For a more
detailed discussion of this topic, we refer the reader to F99 and
Ferraro et al. 2000b, and the references therein.
 
Here, we just note that NGC\,6712 turns out to have, according to Table
2 in F99, the following estimated metallicity in different scales:
$[Fe/H]_{Z85}=-1.01$, $[Fe/H]_{CG97}=-0.88$, and $[M/H]=-0.71$.

\subsection{RGB parameters: metallicity and reddening}
 
A complete set of RGB parameters and metallicity  indicators in the
classical $(V,B-V)$ and in the IR plane has been recently presented by
F99 and Ferraro et al. (2000b), respectively.  These papers also
reported on an independent calibration of RGB-parameters in terms of
the cluster metallicity (both in the CG97  and {\it global} scales).
In particular, F99 presented a system of equations (see Table 4 in F99)
which can be used to simultaneously derive an estimate of metal
abundance (in terms of $[Fe/H]_{CG97}$ and $[M/H]$) and reddening from
the morphology and location of the RGB in the $(V,B-V)$ CMD.  This
system of equations is the equivalent to the so-called SRM method
defined by Sarajedini (1994) in the $(V,V-I)$ plane.

For NGC\,6712 we used the mean ridge line listed in Table 1 to measure
three RGB-parameters:  $(B-V)_g$, defined as the observed RGB colour at
the HB level; the two RGB slopes $S_{2.0}$ and $S_{2.5}$ defined as the
slope of the line connecting the intersection of the RGB and HB with
the points along the RGB located, respectively, 2.0 and 2.5 mag
brighter than the HB.  Assuming the ZAHB level obtained in Section 4
and the mean ridge line listed in Table 1, we obtained a first set of
RGB parameters, namely $(B-V)_g=1.26$, $S_{2.0}=4.69$ and
$S_{2.5}=4.20$ (see F99 for the definition of these parameters).  With
these values as {\it "input"}, the method soon  achieved  convergence
on the following values: $E(B-V)=0.33\pm0.05$,
$[Fe/H]_{CG97}=-0.80\pm0.20$ and $[M/H]=-0.62\pm0.20$.  The errors
associated with the measurements  are conservative estimates of the
global uncertanties, formal errors being much smaller than those
assumed.

The photometric estimate of the cluster metallicity derived with the
technique above fully agrees with previous estimates and confirms that
NGC\,6712 is an intermediate-high metallicity cluster.
  
On the other hand, the value of the reddening obtained with this
procedure appears significantly lower than that listed in the
literature:  $E(B-V)=0.48$, Webbink (1986) and $E(B-V)=0.46$, Harris
(1996).  Past reddening determinations for this cluster, however, are
quite uncertain:
0.48 (Sandage \& Smith, 1966),
0.38 (Kron \& Guetter, 1976),
0.40 (Alcaino, 1977),
0.33 (Martins \& Harvel 1981),
0.42 (Cudworth 1988).  Moreover, Janulis \& Smriglio (1992) confirmed the
existence of differential reddening in the region of NGC\,6712.

\section{Blue Stragglers}

BSS were first observed in the 1950's in the outer region of the GGC M3
(Sandage 1953), while the first sample of BSS in the cores of dense GGC
were discovered with HST (Paresce et al. , 1991) in 47 Tuc.  Since
then, systematic searches for these objects in GGC cores have been
performed from the ground and from  space.  The main result of this
extensive search is that they have been found in all the GGC properly
observed, thus suggesting that BSS are a normal component of the GGC
population (see the catalogue by Sarajedini 1993, and Ferraro,
Bellazzini \& Fusi Pecci 1995).
 
BSS are thought to form via two main mechanisms: (1) the merger of two
stars in a primordial binary system and (2)  stellar collisions (Bailyn
1995, Bailyn \& Pinsonneault 1995).  Many authors (Bailyn 1992, Fusi
Pecci et al. 1992, Ferraro, Bellazzini \& Fusi Pecci 1995) suggested
that BSS in loose clusters might be produced from the coalescence of
primordial binaries, whereas in high density clusters BSS might arise
from stellar collisions (depending on the survival-destruction rate of
primordial binaries).  In particular, collisional BSS can be used as a
diagnostic of the dynamical evolution of GGC (see, for example, the
exceptionally large population of BSS recently found in M80 by Ferraro
et al. 1999b).

From the analysis of the CMD plotted in Figure 4 and Figure 6,
one sees that NGC\,6712 has a quite large population of BSS.
As already said, this population is not affected by field contamination,
since the  BSS sequence is significantly bluer than the Field MS.

On the basis of their location in the $(V,B-V)$ CMD we identified a sample of 
108 BSS. 
The identification numbers, the V,B,R and U magnitudes and the X and Y
coodinates (in pixels [1 px$=0".1$]  with respect to the adopted
cluster center) for the candidate BSS are listed in Table 2. 
Figure 8 shows zoomed CMDs where the BSS candidates are plotted as
large filled circles. As can be seen, all the candidates selected
in the  $(V, B-V)$ CMD lie also in the BSS region in the $(U, U-V)$ CMD. 
Moreover, this latter CMD suggests that a few additional objects might be
considered BSS candidates. We conservatively decided, however, 
to limit our analysis to those stars which are
BSS candidates in both CMDs. In particular, we
intended to avoid regions
of the CMD which can be populated by {\it artificial} stars
resulting from blending effects (see Ferraro et al. 1991, 1992).

It is worth noticing that five objects slightly redder and brighter
that the BSS sequence (plotted as stars in Figure 8) seem to delineate
a track for post-BSS evolution (E-BSS evolved BSS).
  According to their position in the
CMD, they could be BSS which have left the MS to evolve toward the RGB
(BSS in SGB phase).  Candidate BSS in such an evolutionary phase have
been observed in other clusters (see for example the CMD of NGC5053 by
Nemec \& Cohen 1989 or the CMD of the external region of M3, Buonanno
et al. 1994).  We did not, however, include these stars in our sample
as they lie in a region of the CMD where field contamination starts to
be severe. Since we performed only a statistical decontamination of the
observed sample, cluster membership of stars plotted in Figure 6 cannot
be guaranteed, expecially in highly contaminated regions of the CMD.
The selected E-BSS candidate are, nevertheless, worth further
investigation in order to better assess their membership and true
nature and for this reason they are also listed in Table 2.

Even with the conservative selection criterion adopted above, the
number of BSS  found in the central regions of NGC\,6712 is
surprisingly large.  Although in absolute terms, NGC\,6712 has much
fewer BSS than M80 (Ferraro et al. 1999b), it shows a population
comparable with that of NGC\,6388 and NGC2808 (see Sosin et al. 1998)
which are much more massive (a factor $\sim 16$) and concentrated (a
factor  $\sim 100-500$) than NGC\,6712.

In order to quantitatively compare the BSS population across GGC, its
fraction with respect to the total has to be taken into account.  In
doing so, Ferraro et al. (1997, 1999b) defined the specific fraction of
BSS $F_{HB}^{BSS}=N_{BSS}/N_{HB}$ as the number of BSS normalised to
the number of HB stars observed in the same cluster region. In
NGC\,6712 the specific frequency calculated in this was is $\sim 0.69$
similar to that obtained in M3 (Ferraro et al. 1999b) which is again
more massive ( $M_V=-8.75$) and concentrated ($c=2.0$ ) than
NGC\,6712.
  
If one limits the comparison to clusters with similar structure (for
example with comparable concentration  $c\sim1$ and integrated
magnitude), this population appears indeed large. Note that NGC5897
($M_V=-7.27$ and $c=0.95$), for example, has only 34 BSS (Ferraro, Fusi
Pecci \& Buonanno 1992) and $F_{HB}^{BSS}=0.5$.
 
Generally, BSS are more centrally concentrated with respect to the
other stars in the cluster, although there are notable exceptions to
this general rule, such as M\,3 (Ferraro et al. 1993, 1997) and M\,13
(Paltrinieri et al. 1998).  In NGC\,6712 we  compared the radial
distribution of BSS  with respect to RGB and SGB objects (in the range
$17.5<V<20$). The cumulative radial distributions are plotted, as a
function of the projected distance ($r$) from the cluster centre, in
Figure 9, where one sees that BSS (solid line) are more centrally
concentrated than the RGB+SGB sample (dashed line).  The
Kolmogorov-Smirnov test shows that the probability of drawing the two
populations from the same distribution is only $0.1\%$.  The two
distribution are, thus,  different at a high level of significance ($>3
\sigma$).

\section{Blue Objects}

\subsection{A post-AGB bright object, or two?}

Figure 6 shows that a few blue objects are present in the very central
regions of NGC\,6712.  The three faint blue objects ($V>19$ and
$B-V<0.5$), namely star $\#10261$, $\#9774$ and $\#8916$, in our
catalogue, have been discussed in {\it Paper I}.  In particular, star
$\#9774$ is star $S$ of Anderson et al. (1993) and, on the basis of its
position, is the best candidate to the optical counterpart to the
luminous X-ray source detected in this cluster. Star $\#10261$ is a
relatively bright UV star, located a few arcsec away from the X-ray
source, and it is the only object in the HR-field showing a significant
$ H\alpha$ emission (see next Section). For these reasons, in {\it
Paper I} we suggested  it is an additional  promising  interacting binary
candidate.

From the analysis of Figure 6, it is also evident that the bluest
object in the CMD is a bright star (namely $\# 9620$) roughly located
at  the HB level, $V=16.67$ but significantly bluer ($B-V=0.03$) than
the bluest HB  stars ($B-V \sim 0.3$)

Figure 10 shows  the $(U,U-B)$ CMD of all the stars plotted in Figure
6, and star $\# 9620$ still shows as the most UV-bright object in the
field.  The three faint UV stars discussed in {\it Paper I} are plotted
as open triangles.

The position of the UV-bright star in the CMD closely resembles that of
the UV-bright post-AGB star found in M3 (vZ1128, see Buonanno et al.
1994). Such objects are indeed very rare in GGC: only a few post-AGB
stars have been found in GGC due to their short evolutionary lifetime
of $\sim 10^5$ yr (only 0.5 Post-AGB stars are expected to be found in
a typical $\sim 10^5 L_{\odot}$ cluster).

To assess whether the position of star $\# 9620$ is compatible with
that of a post-AGB star, we performed a qualitative comparison with
theoretical models.  A $t\sim 12.5$ Gyr (see Section 8.2) isochrone
with appropriate metallicity ($Z=0.004$) from Bertelli et al. 1994 has
been over-plotted in the CMD of Figure 10, after having been shifted to
fit the main loci, so as to mark the location in this diagramme of the
post-AGB track and the subsequent cooling sequence.  The location of
star $\# 9620$ in the CMD nicely agrees with that predicted by the
models.
 
Note that star $\# 9620$ is located only $\sim 19''$  away from the
cluster centre. Although, in principle, we cannot exclude that this
object is a field star, it is worth  noting that no similar  UV-bright
objects have been found in the sample that we assumed as FIELD (see
Figure 4d).  Moreover, another similar UV-bright object (lying in the
same region of the CMD) has been found in the field labelled as SR-Q1
in Figure 1, at $4.4'$ from the cluster centre (see Figure 4c).  These
objects deserve a more detailed spectroscopic analysis, as proving that
both are cluster members would add further support to the scenario that
NGC\,6712 was much more massive in the past.

\subsection{The $(H\alpha, H\alpha-R)$ CMD}

Some stars in GGC are expected to show $H\alpha$-emission: a few bright
red giants (see Cacciari \& Freeman 1983) or rare objects possibly
connected to planetary nebulae (see Jacoby et al. 1997) or some exotic
objects connected to interacting binaries (as cataclysmic variable and
LMXB; see Cool et al. 1995).  According to Cool et al. 1995, the
$(H\alpha, H\alpha-R)$ CMD can  efficiently serve the purpose of
revealing $H\alpha$-excess objects.  Indeed, in this diagramme stars
with $H\alpha$ in absorption  will have $(H\alpha-R)>0$ while stars
with $H\alpha$ in emission  will feature $(H\alpha-R)<0$.  Typical cool
giants and SGB stars in a cluster will have a weak $H\alpha$ absorption
line ($EW(H\alpha)=1 A$) and, for this reason, they will have
$(H\alpha-R)\sim 0$.

Figure 11 shows the $(H\alpha, H\alpha-R)$ CMD for stars in the HR
field.  The brightest RGB stars are saturated in the long $H\alpha$
exposure, so they are not plotted in that figure.  As expected, the
bulk of the cluster population, constitued as it is by relatively cool
stars, defines a sequence lying in the CMD at $H\alpha-R\sim 0$. The
$H\alpha$-excess star discussed in {\it Paper I} is marked by its
identification number in our catalogue.  Interestingly enough, on the
right of the main sequence  there are a few stars which seem to define
a secondary sequence at $(H\alpha-R)\sim 0.15$ almost parallel to the
main one.  According to their position in this diagramme, they should
have $H\alpha$ in absorption.  These stars are the hottest HB stars and
BSS.  In particular, blue HB stars (with $(B-V)<0.5$ see Figure 6)
which are plotted as large filled circles define  a very narrow
sequence, well defined and well separated from the main one.  The
bright UV-object (star $\#9620$) discussed in the previous section is
plotted in Figure 11 as an open square. According to its position it
turns out to have an $H\alpha$ absorption line of intensity
intermediate between that of giants and blue HB stars.
 
Figure 11 shows that the $(H\alpha, H\alpha-R)$ CMD can be used as an
independent method to select blue HB stars, and can provide a direct
measure of the strength of the $H\alpha$ absorption line in these stars
(see Cool et al. 1995).

\section{Turn-Off point and age}
 
\subsection{Comparison with NGC\,6171: the relative age of NGC\,6712}
 
An accurate determination of  relative ages of GGC provides fundamental
clues to the formation time-scale of the Galactic halo (see Buonanno,
Corsi \& Fusi Pecci 1989, Vandenberg, Stetson \& Bolte 1996, Rosenberg
et al. 1999 -hereafter R99- and reference therein).

Since the SGB-TO region in the CMD plotted in Figure 6 is very well
defined and populated, we can use it to locate  the MS-TO, which is, as
well known, a good age indicator.
 
As mentioned above, the whole CMD of NGC\,6712 closely resembles that
of NGC\,6171, a cluster with similar 
metallicity. In Figure 12 the mean ridge lines of NGC\,6171 (solid
lines) from Ferraro et al. 1991, have been shifted (by $\Delta V=0.65$
and $\Delta (B-V)=0.0$) to match the CMD of NGC\,6712.  The ZAHB level
for NGC\,6171 has been assumed to be $V_{ZAHB}=15.70\pm0.10$ according
to  Table 2 of F99). Figure\,12 shows that the mean loci of NGC\,6121
agree quite nicely with the CMD of NGC\,6712.  Small residual
differences in colour between the mean ridge line and the CMD in the
extreme red region and at the TO region are  probably due to small
residual uncertainties in the colour equation affecting the relatively
old photometry of NGC\,6171.

Note that the lack of any colour shift corresponds exactly  to what is
expected for these two clusters since they have formally the same
reddening (see Ferraro et al. 1991 for NGC\,6171).

To obtain a quantitative estimate of the relative age of the two
clusters we used the so-called {\it vertical method}:  this is  the
most widely used relative-age indicator, based on the TO luminosity
with respect to the HB level (Buonanno, Corsi \& Fusi Pecci 1989).
 
From Table 1, the TO level in NGC\,6712 is located at $V_{TO}=19.82\pm
0.10$. Using this value and the $V_{ZAHB}$  computed in Section 4, we
derived $\Delta V_{TO}^{HB}=3.50\pm 0.15$ mag.  For NGC\,6171, assuming
$V_{TO}^{NGC\,6171}=19.20\pm0.10$ (Ferraro et al. 1991) and
$V_{ZAHB}^{NGC\,6171}=15.70\pm0.10$ (from F99), we found $\Delta
V_{TO}^{HB}=3.50\pm 0.15$ mag.
 
From  this comparison, we can conclude that NGC\,6712 is nearly coeval
with NGC\,6171.  On the other hand, NGC\,6171 has been found (R99) to
be coeval with  other clusters belonging to the same metallicity class
(NGC288, NGC\,6218, NGC\,6362, NGC\,6723).  Thus, we can safely
conclude that NGC\,6712 does not show any significant difference in age
with respect to the bulk of the intermediate metal-rich population of
GGC.

\subsection{Cluster distance and absolute age}

The evaluation of absolute ages requires the knowledge (or the implicit
assumption) of the distance modulus for the cluster. Different loci in
the CMD have been assumed in the past  as standard candles (HB - see
F99, MS - see Gratton et al. 1997, WD - see Renzini et al. 1996) to
determine cluster distances. F99 used the HB and, in particular, the
ZAHB as a reference candle and derived a new homogeneous distance scale
for a sample of 64 GGC. Here we use the same procedure to derive the
distance to NGC\,6712.

The Straniero Chieffi \& Limongi (1997) theoretical HB models were used
to derive the absolute magnitude of the ZAHB as a function of the
metallicity (see eq. 4 in F99). From this relation and assuming
$[Fe/H]_{CG97}=-0.8$ (section 5.2), we get $M_V^{ZAHB}=0.75$. Assuming
$V_{ZAHB}=16.32$ (see Section 4), we obtain an apparent distance
modulus $(m-M)_V=15.57$ and finally (considering $E(B-V)=0.33$, see
Section 5.2) a true modulus of $(m-M)_0=14.55$, which corresponds to a
distance of $\sim 8.1$ Kpc.

This value turns out to be significantly larger than that obtained by
C88 ($6.5$ Kpc). The origin of such a large discrepancy is due to many
differing assumptions between C88 and this paper.  First of all, the
reddening adopted here ($E(B-V)=0.33$) is quite different from that
used by  C88 ($E(B-V)=0.42$).  Second, the assumption of the absolute
level of the HB is different:  C88 used $M_V^{HB}=0.85$ and we used
$M_V^{HB}=0.75$ and finally there is a small difference (0.07 mag) in
the observed value of the HB level (C88 used $V_{HB}=16.25$).

In addition, C88 adopted the relation $A_V=3.2E(B-V)$ instead of the
canonical $R_V=3.1$. This (or even higher) value of $R_V$ can be
reasonable for NGC\,6712 since it is in a highly contaminated region.
Although the different assumption for $R_V$ between us and C88 produces
only a small effect($\sim 0.04$ mag), the impact of a different $R$
coefficient on the determination of the true distance modulus of
NGC\,6712 should be kept in mind.  At the moment, however, the main
source of uncertainty in the distance error budget  is  the reddening:
the assumed error  ($\pm 0.05$), in fact, produces an uncertainty of
$\sim 0.15$ mag in the distance modulus and an error of $\sim 6\%$
($\sim 0.5$ Kpc) in the true distance.  Taking into account this and
the uncertainties on the other parameters concurring to the
determination of the distance, we finally conclude that the  global
uncertainty cannot be less than $10\%$ and assume $d=8\pm1$ Kpc as a
final conservative estimate of the distance to NGC\,6712.

From the distance modulus obtained above and $V_{TO}=19.82\pm0.10$
(from Table 1), we get $M_V^{TO}=4.25\pm0.2$.  With this value and the
metallicity estimate ($[Fe/H]_{CG97}=-0.8$) obtained in Section 5.2, we
can derive the age for NGC\,6712 using theoretical relations linking
these observables to the cluster age.  In doing so, we used the
relations recently obtained by R99 (see their Appendix A), for three
sets of theoretical models:  SCL97, Cassisi et al. (1998) and
VandenBerg et al. (2000) and obtained $t=12.6$,  $12.9$ and $11.5$ Gyr,
for SCL97, Cassisi et al. (1998) and VandenBerg et al. (2000),
respectively.  Taking into account that the global uncertainty of the
method can hardly be better than $15\%$, we conservatively assume
$t=12\pm2$ Gyr as the absolute age of NGC\,6712.

\section{Summary and Conclusions}

In this paper we have presented extensive, high resolution, $UBVR$
photometry of the galactic globular cluster NGC\,6712, obtained with
the ESO-Very Large Telescope.  This data set gives us the opportunity
to investigate the evolved stellar population of the cluster, yielding
new constraints on  its metallicity, age and  structural parameters.
In particular we find the following results:

\begin{itemize}
 
\item{\it Metallicity and Reddening:} The location and morphology of
the RGB  in the (V,B-V)-CMD, compared with a large set of reference
clusters, fully confirm that NGC\,6712 is an intermediate metal-rich
cluster. Using the method described in F99 we get $[M/H]=0.62 \pm 0.2$,
$[Fe/H]_{CG97}=-0.80\pm0.2$.  The method also yields an estimate of the
reddening which  turns out to be $E(B-V)=0.33\pm0.05$, a value
significantly lower than previously thought.

\item{\it Distance:} Using the theoretical level of the ZAHB (from
SCL97) as a reference candle, we derived $(m-M)_V=15.57$ and a true
distance modulus $(m-M)_0=14.55$ which corresponds to a distance of
about $\sim 8$ Kpc. This value is significantly higher than that
obtained by C88 (who found 6.5 Kpc), mainly because of the new reddening
determination.
 
\item{\it Age:} Coupling the apparent luminosity of the  ZAHB,
$V_{ZAHB}=16.32\pm0.05$ with $V_{TO}=19.82\pm0.10$, we get $\Delta
V_{TO}^{HB}=3.5\pm0.1$, a value fully compatible with that obtained in
other clusters.  This suggests that NGC\,6712 is nearly coeval, at
least, with other clusters with similar metallicity.  An average
absolute age of $t=12\pm 2 Gyr$ has been derived from comparison with
the theoretical models.
  
\item{\it BSS:} The  cluster reveals a large population of BSS: we
selected a sample of 108 BSS from $(V,B-V)$ and $(U,U-V)$ CMD. This
number is surprisingly large when compared with the BSS  population of
clusters with similar mass and central concentration.
 
\item{\it Post-AGB star:} We discovered the presence of a bright
UV-bright object in the core, whose position in the CMD closely matches
that of a star evolving in the post-AGB phase. A second objects with
similar characteristics has been located further out in the cluster at
$\sim 4\farcm4$ from the centre.
 
\end{itemize}

The overall scenario emerging when we combine some of the results
listed above with those already presented in the two previous papers of
this series ({\it Paper I} and {\it II}), is quite exciting.
 
NGC\,6712 is experiencing a severe interaction with the disk and the
bulge of the Galaxy, as suggested by its orbit (Dauphole et al.
1996).  The existence (presented in De Marchi et al. 1999, and
confirmed by {\it Paper II}) of an {\it inverted} mass function due to
a severe depletion of low mass stars in the MS luminosity function,
probably stripped away by the tidal force of the Galaxy, fully supports
this scenario.
 
Takahashi \& Portegies Zwart (2000), using N-body simulations,
interpreted this rare feature as a clear signature of severe mass loss
and suggested that NGC\,6712 might be only the remnant "core"  of a
once much more massive cluster, that  has lost most (up to $\sim 99
\%$) of its initial mass through the interaction with the Galaxy.  Thus
the hypothesis  that NGC\,6712 might have been much more massive (and,
possibly, much more concentrated) in the past, gained  support from a
theoretical point of view.

Now, this scenario finds additional support from a variety of
observational facts.  The presence of a LMXB and our recent discovery
({\it Paper I}) of an additional promising interacting binary candidate
located a few arcsec away from the optical counterpart to the LMXB,
suggest that strong stellar interactions might have  occurred at some
remote stage of the cluster evolution.  Moreover, the large BSS
population discovered in this paper adds additional support to the fact
that star collisions might have occurred in the past,  generating most
(or part) of the observed BSS. At that time, NGC\,6712 probably was a
massive and concentrated cluster and collisional BSS (and other exotic
objects such as interacting binaries) formed copiously via dynamical
collisions. Then, these stars have migrated towards the centre, because
of mass segregation, where we now see them.  The continued action of
tidal stripping and disk shocking has removed most of the cluster mass,
driving it towards dissolution.  What we now observe is nothing but
the  remnant core of a disrupting cluster and its population of
peculiar objects, which are otherwise totally unexpected for its actual
mass.

If this scenario is further confirmed, our observations would have
served the purpose of reconstructing the stormy history of NGC\,6712
and possibly, for the very first time, to characterize the effects of
the disk shocking on the evolution of  stars in a globular cluster.

\acknowledgments

This research was partially supported by the {\it Agenzia Spaziale
Italiana} (ASI).  The financial support of  the {\it Ministero della
Universit\`a e della Ricerca Scientifica e Tecnologica} (MURST) to the
project {\it Stellar Dynamics and Stellar Evolution in Globular
Clusters} is kindly acknowledged.  F. R. F. gratefully acknowledges the
hospitality of the {\it Visitor Program} during his stay at ESO, when
most of this work has been carried out.  The data presented in this
paper were obtained as part of an ESO Service Mode programme.

\clearpage

\begin{deluxetable}{cccccccc}
\tablewidth{\textwidth}
\label{}
\tablecaption{RGB-SGB Mean ridge line}
\tablehead{
\colhead{B-V} &
\colhead{V} & 
 &
\colhead{B-V} &
\colhead{V} & 
&
\colhead{B-V}&
\colhead{V}
}
\startdata
     2.110 &    13.510 &  &   1.214 &    16.700 & &    0.932 &     19.370 \nl
    2.048 &    13.540 &   &  1.195 &    16.900 &  &  0.910 &     19.420 \nl
    1.978 &    13.570 &   &  1.176 &    17.100 &  &   0.888 &     19.460 \nl
    1.909 &    13.700 &   &  1.155 &    17.300 &  &   0.875 &     19.490 \nl
    1.824 &    13.900 &   &  1.138 &    17.500 &  &   0.861 &     19.530 \nl
    1.761 &    14.100 &   &  1.127 &    17.700 &  &   0.845 &     19.580 \nl
    1.691 &    14.300 &   &  1.115 &    17.900 &  &   0.832 &     19.650 \nl
    1.640 &    14.500 &   &  1.103 &    18.100 &  &   0.826 &     19.730 \nl
    1.580 &    14.700 &   &  1.096 &    18.300 &  &   0.822 &     19.820 \nl
    1.521 &    14.900 &   &  1.085 &    18.500 &  &   0.824 &     19.900 \nl
    1.478 &    15.100 &   &  1.079 &    18.700 &  &   0.826 &     19.990 \nl
    1.435 &    15.300 &   &  1.061 &    18.930 &  &   0.829 &     20.090 \nl
    1.392 &    15.500 &   &  1.048 &    19.110 &  &   0.835 &     20.200 \nl
    1.356 &    15.700 &   &  1.023 &    19.220 &  &   0.842 &     20.300 \nl
    1.325 &    15.900 &   &  1.008 &    19.250 &  &   0.856 &     20.460 \nl
    1.288 &    16.100 &   &  0.995 &    19.280 &  &   0.868 &     20.580 \nl
    1.261 &    16.300 &   &  0.973 &    19.310 & &    0.894 &     20.810 \nl
    1.240 &    16.500 &   &  0.953 &    19.330 & &    0.917 &     21.020 \nl
\enddata
\end{deluxetable}

\clearpage

\begin{deluxetable}{ccccccccc}
\tablewidth{\textwidth}
\label{}
\tablecaption{BSS candidates in NGC6712}
\tablehead{
&
\colhead{\#} &
\colhead{ID} &
\colhead{V} & 
\colhead{B} &
\colhead{R} & 
\colhead{U} &
\colhead{X} & 
\colhead{Y}  
 }
\startdata
  &   BSS  1 &   111 & 19.45 & 20.18 & 18.95 & 20.46 & -400.973 & -998.374 \nl
  &   BSS  2 &   144 & 19.36 & 20.10 & 18.85 & 20.20 &  804.297 & -980.577 \nl
  &   BSS  3 &   262 & 18.56 & 19.27 & 18.08 & 19.58 &  728.953 & -916.481 \nl
  &   BSS  4 &   540 & 17.75 & 18.33 & 17.35 & 18.65 &  792.174 & -791.430 \nl
  &   BSS  5 &   670 & 18.02 & 18.68 & 17.57 & 19.05 & -322.614 & -743.712 \nl
  &   BSS  6 &  1155 & 18.66 & 19.31 & 18.20 & 19.56 & -196.280 & -593.166 \nl
  &   BSS  7 &  1177 & 18.57 & 19.24 & 18.11 & 19.48 & -117.403 & -598.961 \nl
  &   BSS  8 &  1242 & 18.79 & 19.36 & 18.39 & 19.71 &  -45.187 & -580.513 \nl
  &   BSS  9 &  1254 & 18.51 & 19.25 & 17.99 & 19.32 &  532.380 & -579.290 \nl
  &   BSS 10 &  1456 & 18.93 & 19.62 & 18.47 & 19.84 &   20.270 & -526.844 \nl
  &   BSS 11 &  1504 & 18.52 & 19.11 & 18.12 & 19.52 &  303.193 & -523.704 \nl
  &   BSS 12 &  1568 & 18.36 & 18.93 & 17.99 & 19.24 & -398.783 & -481.762 \nl
  &   BSS 13 &  1571 & 19.23 & 19.96 & 18.76 & 20.15 & -426.108 & -486.179 \nl
  &   BSS 14 &  1742 & 18.76 & 19.43 & 18.28 & 19.65 &  -76.106 & -473.851 \nl
  &   BSS 15 &  1850 & 18.58 & 19.35 & 18.06 & 19.63 & -638.202 & -449.283 \nl
  &   BSS 16 &  1903 & 19.12 & 19.77 & 18.64 & 19.95 &  112.571 & -438.082 \nl
  &   BSS 17 &  1906 & 18.77 & 19.54 & 18.22 & 19.76 &  120.261 & -427.270 \nl
  &   BSS 18 &  1920 & 19.48 & 20.14 & 19.03 & 20.44 & -190.172 & -436.096 \nl
  &   BSS 19 &  1930 & 19.00 & 19.66 & 18.54 & 19.92 & -888.934 & -431.771 \nl
  &   BSS 20 &  2042 & 18.81 & 19.57 & 18.32 & 19.79 &  331.887 & -415.190 \nl
 &   BSS 21 &  2428 & 19.39 & 20.14 & 18.88 & 20.33 &  528.692 & -351.818 \nl
  &   BSS 22 &  2544 & 19.09 & 19.78 & 18.62 & 20.02 &  170.663 & -335.835 \nl
  &   BSS 23 &  2546 & 19.36 & 20.05 & 18.91 & 20.23 &  162.471 & -335.651 \nl
  &   BSS 24 &  2547 & 19.08 & 19.71 & 18.65 & 19.93 &  152.264 & -330.614 \nl
  &   BSS 25 &  2623 & 18.79 & 19.49 & 18.30 & 19.67 & -330.428 & -300.632 \nl
  &   BSS 26 &  2749 & 19.23 & 19.93 & 18.76 & 20.11 & -305.596 & -305.598 \nl
  &   BSS 27 &  2965 & 19.25 & 19.96 & 18.78 & 20.10 & -293.869 & -268.150 \nl
  &   BSS 28 &  2978 & 18.56 & 19.19 & 18.14 & 19.48 &  -45.357 & -269.560 \nl
  &   BSS 29 &  3171 & 19.10 & 19.82 & 18.61 & 20.01 &  193.588 & -237.935 \nl
  &   BSS 30 &  3253 & 18.72 & 19.30 & 18.31 & 19.56 &  244.559 & -231.358 \nl
  &   BSS 31 &  3434 & 18.60 & 19.21 & 18.19 & 19.46 & -799.539 & -201.215 \nl
  &   BSS 32 &  3501 & 18.79 & 19.47 & 18.33 & 19.68 & -805.423 & -188.600 \nl
  &   BSS 33 &  3568 & 17.62 & 18.23 & 17.21 & 18.62 &  166.820 & -167.683 \nl
  &   BSS 34 &  3684 & 18.69 & 19.49 & 18.14 & 19.67 &  208.239 & -157.790 \nl
  &   BSS 35 &  3814 & 18.60 & 19.36 & 18.08 & 19.57 &  -59.101 & -139.850 \nl
  &   BSS 36 &  4087 & 19.38 & 20.15 & 18.84 & 20.23 & -319.318 & -101.848 \nl
  &   BSS 37 &  4157 & 17.68 & 18.13 & 17.36 & 18.56 & -790.025 &  -91.930 \nl
  &   BSS 38 &  4239 & 19.39 & 20.13 & 18.89 & 20.29 & -492.933 &  -77.139 \nl
  &   BSS 39 &  4329 & 19.29 & 20.07 & 18.80 & 20.22 &  543.722 &  -30.792 \nl
  &   BSS 40 &  4332 & 18.32 & 19.05 & 17.82 & 19.23 &  633.899 &  -62.242 \nl
  &   BSS 41 &  4337 & 18.61 & 19.17 & 18.22 & 19.46 &  739.292 &  -51.804 \nl
  &   BSS 42 &  4350 & 19.32 & 19.97 & 18.88 & 20.11 & -764.300 &  -56.442 \nl
  &   BSS 43 &  4431 & 17.95 & 18.40 & 17.61 & 18.75 &  186.752 &  -46.434 \nl
  &   BSS 44 &  4463 & 18.87 & 19.49 & 18.45 & 19.69 & -411.709 &  -32.642 \nl
  &   BSS 45 &  4469 & 18.63 & 19.23 & 18.22 & 19.49 &  295.844 &  -36.940 \nl
  &   BSS 46 &  4470 & 19.28 & 19.90 & 18.84 & 20.08 &  287.043 &  -32.371 \nl
  &   BSS 47 &  4528 & 18.87 & 19.60 & 18.36 & 19.82 &  -70.825 &  -32.440 \nl
  &   BSS 48 &  4572 & 19.27 & 20.05 & 18.74 & 20.34 & -150.785 &  -25.653 \nl
  &   BSS 49 &  4723 & 19.50 & 20.23 & 18.97 & 20.39 & -325.664 &   -1.551 \nl
  &   BSS 50 &  4845 & 18.58 & 19.34 & 18.07 & 19.75 &  -82.528 &   19.827 \nl
  &   BSS 51 &  5111 & 19.03 & 19.81 & 18.49 & 19.99 & -835.504 &   58.053 \nl
  &   BSS 52 &  5181 & 18.39 & 18.99 & 17.99 & 19.32 &  451.931 &   80.364 \nl
  &   BSS 53 &  5198 & 18.61 & 19.34 & 18.10 & 19.61 &  -56.833 &   69.893 \nl
  &   BSS 54 &  5364 & 19.28 & 19.95 & 18.83 & 20.20 & -198.830 &  108.252 \nl
  &   BSS 55 &  5514 & 19.01 & 19.82 & 18.47 & 20.03 & -336.323 &  121.265 \nl
  &   BSS 56 &  5631 & 18.32 & 19.00 & 17.85 & 19.29 & -404.621 &  146.077 \nl
  &   BSS 57 &  5712 & 19.13 & 19.80 & 18.65 & 19.97 &  329.344 &  150.904 \nl
  &   BSS 58 &  5816 & 18.52 & 19.13 & 18.09 & 19.43 &  172.368 &  179.946 \nl
  &   BSS 59 &  5938 & 17.85 & 18.46 & 17.43 & 18.90 &   50.804 &  191.088 \nl
  &   BSS 60 &  6003 & 18.04 & 18.67 & 17.60 & 18.89 &  889.243 &  202.474 \nl
  &   BSS 61 &  6014 & 17.58 & 18.07 & 17.24 & 18.50 &  621.351 &  212.900 \nl
  &   BSS 62 &  6243 & 18.32 & 19.05 & 17.95 & 19.42 &  324.106 &  250.420 \nl
  &   BSS 63 &  6244 & 19.08 & 19.77 & 18.63 & 19.95 &  317.493 &  259.019 \nl
  &   BSS 64 &  6433 & 18.91 & 19.73 & 18.36 & 19.97 &   38.265 &  280.186 \nl
  &   BSS 65 &  6556 & 19.19 & 19.97 & 18.65 & 20.11 & 1065.295 &  282.922 \nl
  &   BSS 66 &  6594 & 18.65 & 19.29 & 18.21 & 19.48 & -947.905 &  292.308 \nl
  &   BSS 67 &  6671 & 18.87 & 19.54 & 18.41 & 19.79 &  149.435 &  311.900 \nl
  &   BSS 68 &  6699 & 18.96 & 19.77 & 18.42 & 19.96 & -267.150 &  322.899 \nl
  &   BSS 69 &  6701 & 18.99 & 19.78 & 18.44 & 19.99 & -282.224 &  317.300 \nl
  &   BSS 70 &  6797 & 18.98 & 19.74 & 18.52 & 19.96 &  114.881 &  326.295 \nl
  &   BSS 71 &  6817 & 19.20 & 19.95 & 18.67 & 20.15 &  472.171 &  323.492 \nl
  &   BSS 72 &  7298 & 18.76 & 19.42 & 18.31 & 19.66 & -325.091 &  413.188 \nl
  &   BSS 73 &  7305 & 18.68 & 19.30 & 18.25 & 19.51 &  259.097 &  416.515 \nl
  &   BSS 74 &  7506 & 18.43 & 19.16 & 17.92 & 19.41 &   40.085 &  454.635 \nl
  &   BSS 75 &  7563 & 19.00 & 19.74 & 18.50 & 19.90 & -571.343 &  465.771 \nl
  &   BSS 76 &  7963 & 19.08 & 19.86 & 18.54 & 20.01 &  470.976 &  541.285 \nl
  &   BSS 77 &  8036 & 18.04 & 18.64 & 17.62 & 18.87 &  428.476 &  568.233 \nl
  &   BSS 78 &  8193 & 18.19 & 18.81 & 17.74 & 19.07 &  306.234 &  601.330 \nl
  &   BSS 79 &  8918 & 18.20 & 18.89 & 17.73 & 19.10 &  186.325 &  809.361 \nl
  &   BSS 80 &  9113 & 19.04 & 19.82 & 18.50 & 19.99 &  479.085 &  879.267 \nl
  &   BSS 81 &  9115 & 19.23 & 19.88 & 18.78 & 20.07 & -528.758 &  889.671 \nl
  &   BSS 82 &  9294 & 19.24 & 20.02 & 18.71 & 20.09 &  288.298 &  954.655 \nl
  &   BSS 83 &  9297 & 19.24 & 19.99 & 18.73 & 20.30 &  564.001 &  953.698 \nl
  &   BSS 84 &  9317 & 19.03 & 19.67 & 18.58 & 19.85 &   -1.994 &  964.133 \nl
  &   BSS 85 &  9589 & 19.15 & 19.87 & 18.66 & 20.08 & -316.575 & -115.017 \nl
  &   BSS 86 &  9599 & 18.92 & 19.67 & 18.47 & 19.88 &   17.417 & -119.848 \nl
  &   BSS 87 &  9618 & 18.34 & 19.05 & 17.84 & 19.30 &  -91.522 & -110.438 \nl
  &   BSS 88 &  9688 & 18.03 & 18.63 & 17.60 & 18.96 &   11.660 &  -63.064 \nl
  &   BSS 89 &  9723 & 19.23 & 19.92 & 18.73 & 20.15 & -261.537 &  -51.061 \nl
  &   BSS 90 &  9740 & 18.28 & 19.00 & 17.80 & 19.32 & -120.423 &  -35.079 \nl
  &   BSS 91 &  9763 & 18.31 & 18.96 & 17.86 & 19.23 &  -18.104 &  -27.268 \nl
  &   BSS 92 &  9765 & 19.16 & 19.92 & 18.67 & 20.11 &  -19.759 &  -21.688 \nl
  &   BSS 93 &  9797 & 18.18 & 18.71 & 17.83 & 19.25 & -104.733 &  -15.695 \nl
  &   BSS 94 &  9798 & 18.59 & 19.29 & 18.12 & 19.59 & -115.067 &  -13.541 \nl
  &   BSS 95 &  9909 & 18.46 & 19.10 & 18.02 & 19.40 & -185.532 &   66.507 \nl
  &   BSS 96 &  9955 & 17.59 & 18.27 & 17.12 & 18.61 &   40.322 &   84.100 \nl
  &   BSS 97 &  9961 & 19.13 & 19.89 & 18.59 & 20.13 &  177.678 &   71.732 \nl
  &   BSS 98 & 10002 & 18.69 & 19.36 & 18.25 & 19.65 &   15.869 &  106.062 \nl
  &   BSS 99 & 10014 & 18.31 & 18.86 & 17.93 & 19.22 &   77.432 &  115.310 \nl
  &   BSS100 & 10046 & 18.77 & 19.57 & 18.29 & 19.83 & -233.953 &  123.758 \nl
  &   BSS101 & 10053 & 19.22 & 19.92 & 18.72 & 20.14 & -228.244 &  132.875 \nl
  &   BSS102 & 10092 & 17.48 & 18.12 & 17.04 & 18.60 &   16.717 &  167.770 \nl
  &   BSS103 & 10159 & 18.58 & 19.33 & 18.13 & 19.58 &  108.702 &  202.048 \nl
  &   BSS104 & 10226 & 18.40 & 19.04 & 18.05 & 19.05 & -194.530 & -178.189 \nl
  &   BSS105 & 10277 & 19.14 & 19.83 & 18.66 & 20.07 &  -14.713 &  -80.536 \nl
  &   BSS106 & 10286 & 19.14 & 19.87 & 18.67 & 20.13 & -220.658 &  -49.144 \nl
  &   BSS107 & 10365 & 19.02 & 19.76 & 18.29 & 20.04 &  -46.893 &  -75.749 \nl
  &   BSS108 & 10376 & 18.50 & 19.14 & 18.04 & 19.46 &   75.401 &  -18.944 \nl
   &         &        &       &        &    &       &     & \nl
  &   E-BSS  1 &   816 & 17.31 & 18.16 & 16.77 & 18.46 &    6.936 & -685.088 \nl
  &   E-BSS  2 &  1692 & 17.33 & 18.18 & 16.82 & 18.60 &  477.852 & -483.990 \nl
  &   E-BSS  3 &  2533 & 17.14 & 17.92 & 16.64 & 18.39 & -142.159 & -313.275 \nl
  &   E-BSS  4 &  3437 & 17.19 & 17.94 & 16.68 & 18.37 &  768.774 & -200.574 \nl
  &   E-BSS  5 &  3442 & 17.27 & 18.11 & 16.75 & 18.51 & -840.575 & -188.349 \nl
\enddata
\end{deluxetable}

\clearpage
\centerline { FIGURE CAPTION}

Fig. 1 ~~~Location of the five FORS1 fields observed in
NGC\,6712.  Four fields were observed at Standard Resolution (SR):
Q1,Q2,Q3,Q4 and one at High Resolution (HR), respectively.  The HR
field is  roughly centered on the cluster centre.  The cluster centre
is located at the origin of the coordinates.

\smallskip

Fig. 2 ~~~Calibration curves for the U, B, V, R filters.

\smallskip

Fig. 3 ~~~Internal photometric errors for the HR field. The
rms values of the frame-to-frame scatter are plotted versus the U, B,
V, R magnitude of each star.  

\smallskip

Fig. 4 ~~~The $V$, $B-V$ colour magnitude diagramme for the
whole sample of stars observed in NGC\,6712. In  Panel (a) only stars
lieing in the FORS-HR field are plotted, while Panel (b),(c) and (d)
show stars in annuli at increasing distance from the cluster centre.

\smallskip

Fig. 5 ~~~The observed radial density profile for the
evolved stellar population ({\it filled circles}). All stars plotted in
Figure 4 with  $V<20$ have been counted. The stellar density has been
scaled to match the density profile of unevolved MS stars
with $19.5<R<20.5$ (corresponding to $\sim 0.75 M_{\odot}$) for $r>2'$
({\it large open squares}) see also Figure 8 in {\it Paper II}).  The
dotted horizontal line represents the density of $ 0.75 M_{\odot}$
star measured in the control field at $42'$ from the cluster centre.
The distance $r=5'$ from the cluster centre is indicated by  a vertical
arrow.
 
\smallskip
 
Fig. 6 ~~~The $V$, $B-V$ colour magnitude diagramme for
stars observed in the HR field after the statistical decontamination
from field stars (see text)

\smallskip

Fig. 7 ~~~HB stars in NGC\,6712. The solid heavy line is
the $V_{ZAHB}$ level, determined by F99. Seven RR Lyrae variables are
plotted as large filled triangles.

\smallskip

Fig. 8 ~~~Blue Stragglers Star candidates {\it (large
filled circles)} in NGC\,6712.  The zoomed CMD in the BSS region in the
$(V,B-V)$ {\it (panel a)} and $(U,U-V)$ {\it (panel b)}, plane.
The 5 objects plotted as {\it stars} are suspected to be evolved (E-BSS)
{\it i.e.} BSS in the SGB, which have left the MS to evolve toward the RGB.
 
\smallskip

Fig. 9 ~~~Cumulative radial distribution ($\phi$) for BSS
(heavy solid line) compared to the RGB-SGB (dashed line) as a function
of their projected distance ($r$) from the cluster centre in arcmin.

\smallskip
 
Fig. 10 ~~~$U,U-B$ CMD for NGC\,6712 from FORS1-HR images.
The bright blue object $\#9620$ is plotted as a {\it large filled
triangle}.  The three faint UV stars (plotted as {\it open triangles})
are discussed in {\it Paper I}. An isochrone from Bertelli et al.
(1994) is also plotted for reference.  As can be seen, the bright blue
object is located along the post-AGB cooling sequence.

\smallskip
 
Fig. 11 ~~~$(H\alpha, H\alpha-R)$ CMD for NGC\,6712 from
FORS1-HR images. The $H\alpha$ excess star  $\#10261$ (plotted as a
{\it large filled triangle}) is the star discussed in {\it Paper I}.
Blue HB stars (with $B-V<0.5$) are plotted as large filled circles, BSS
as selected in Section 6 are plotted as large asterisks.  The small
open square is the bright blue object discussed in Section 7.1.

\smallskip

Fig. 12 ~~~The mean ridge line of NGC\,6171 shifted
[$\Delta V=0.62, \Delta (B-V)=0.0$] to match the CMD of NGC\,6712.

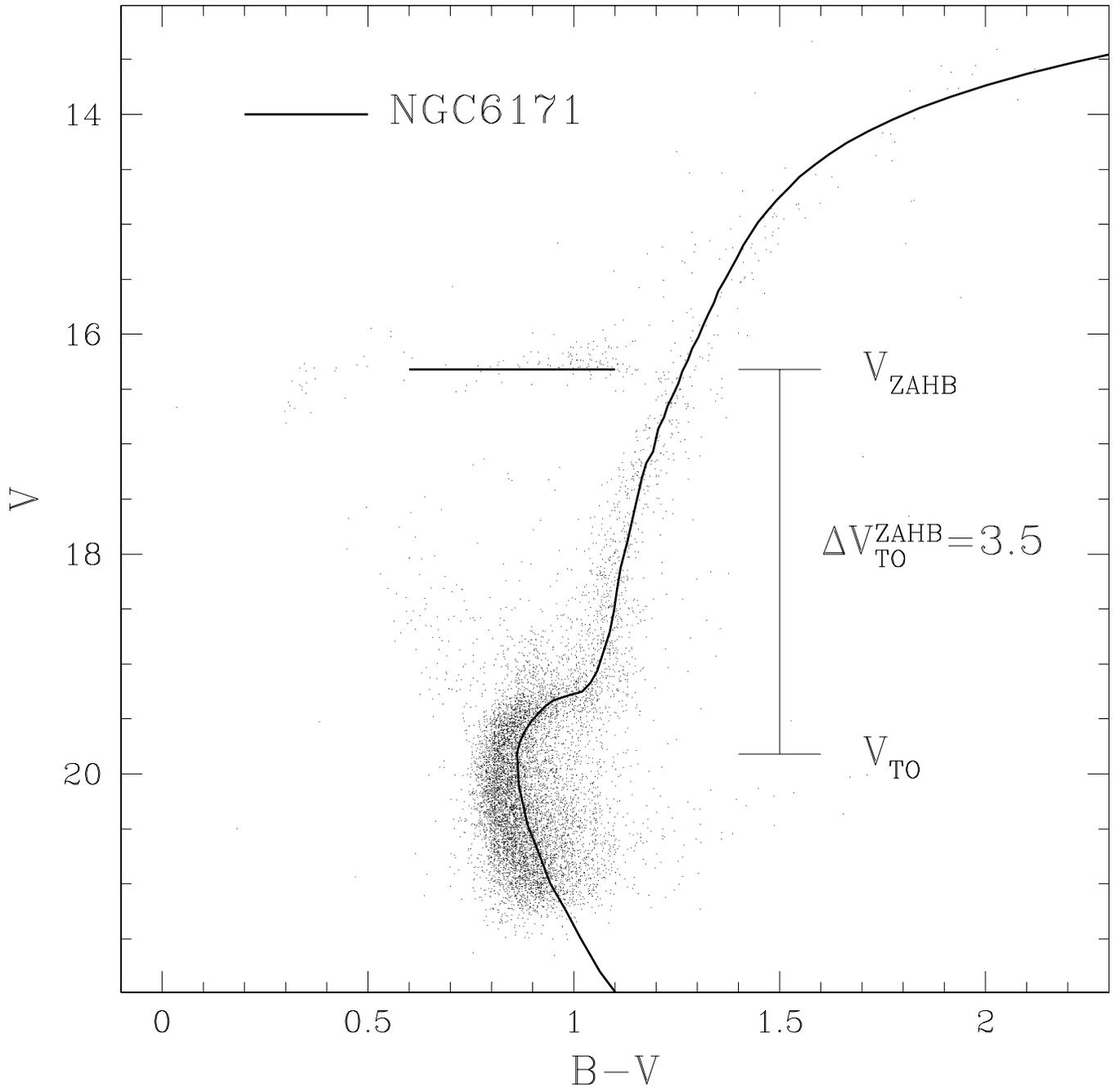
\end{document}